\documentclass[slac_one]{revtex4}
\usepackage{graphicx}
\usepackage{fancyhdr}
\begin{document}

\title{Electron and Photon Identification Performance in ATLAS} %% Paper title goes here


\author{H.J. Kim (for the ATLAS collaboration)}
\affiliation{The University of Texas at Arlington, Arlington, TX 76019, USA}

\begin{abstract}
The understanding of the reconstruction and calibration of electrons and photons is one of the key steps at the start-up of data-taking with ATLAS~\cite{phytdr} at the LHC (Large Hadron Collider). The calorimeter cells are electronically calibrated before being clustered. Corrections to local position and energy measurements are applied to take into account the calorimeter geometry. Finally, longitudinal weights are applied to correct for energy loss upstream of the calorimeter. As a last step the $Z \to ee$ events will be used for in-situ calibration using the $Z$ boson mass. The electron identification is based on the shower shape in the calorimeter and relies heavily on the tracker and combined tracker/calorimeter information to achieve the required rejection of 10$^5$ against QCD jets for a reasonably clean inclusive electron spectrum above $20-25$ GeV. For photon identification, in addition to the shower shape in the calorimeter, recovery of photon conversions is an essential ingredient given the large amount of material in the inner tracker. The electron and photon identification methods (cuts and multivariate analysis) will be discussed.
\end{abstract}

\maketitle

\thispagestyle{fancy}


\section{THE ATLAS ELECTROMAGNETIC CALORIMETER}

The ATLAS electromagnetic (EM) calorimeter is a lead/liquid argon sampling calorimeter with accordion shaped electrodes and absorbers interleaved. The calorimeter is divided in two half barrel cylinders covering the pseudo-rapidity range $|\eta| \leq 1.475$, housed in a single cryostat and two endcap detector (covering $1.375 \leq |\eta| \leq 3.2$) housed in two separate endcap cryostats. Its accordion structure provides complete $\phi$ symmetry without azimuthal cracks. The total thickness of the calorimeter is greater than 22 radiation lengths ($X_{0}$) in the barrel and 24$X_{0}$ in the endcaps. The EM calorimeter is highly segmented with a 3-fold granularity in depth and $\eta \times \phi$ granularity of $0.0003 \times 0.1$, $0.025 \times 0.025$, and $0.05 \times 0.025$, respectively in the front, middle and back compartment. A pre-sampler with a fine granularity in $\eta$ ($\Delta \eta = 0.025$) is located before the cryostat and the coil, enabling to correct for the corresponding dead material effects. More details on the ATLAS detector can be found in~\cite{caltdr}.

\section{ELECTROMAGNETIC CALIBRATION} % Section title should be in all capitals.
The energy measurement in the calorimeter cells is the starting point of the reconstruction of electrons and photons. The construction of cell clusters is based on two algorithms, fixed-size window clusters for photons and topological clusters for electrons. The fixed-size algorithm starts by choosing a seed cell in the middle layer of EM calorimeter and then varies the position of a window to maximize the total energy contained in it. For the topological cluster, cells are chosen as seeds if their energy is above a given threshold. Since the material in front and the segmentation of the calorimeter affect the measured energy and position of EM clusters, position and energy corrections are applied at the cluster level. Due to the finite granularity of the detector, the difference between the true and the computed shower barycenter, as a function of the $\eta$ position inside the cell, has a typical S-shape.  The cluster position in $\phi$ is determined from the energy barycenter in the second sampling. The measurement of $\phi$ is biased by an offset due to the accordion shape and depends on the distance to the folds of the accordion. The energy of a cluster is obtained by $E_{rec} = \lambda (b +  \omega_{0} E_{0} + E_{1} + E_{2} + \omega_{3} E_{3})$, where  $E_{0}$, $E_{1}$, $E_{2}$ and $E_{3}$ are the energies in the pre-sampler and the three layers of calorimeter. The offset term $b$ corrects for upstream energy loss before pre-sampler. The parameters $\lambda$, $b$, $\omega_{0}$, and $\omega_{3}$, called longitudinal weights, are calculated by a $\chi^2$ minimization of  $(E_{true} ~ - ~ E_{rec})^2 /\sigma(E_{true})^2$ using Monte Carlo single particle samples.

\begin{figure*}[t]
\centering
\includegraphics[width=65mm]{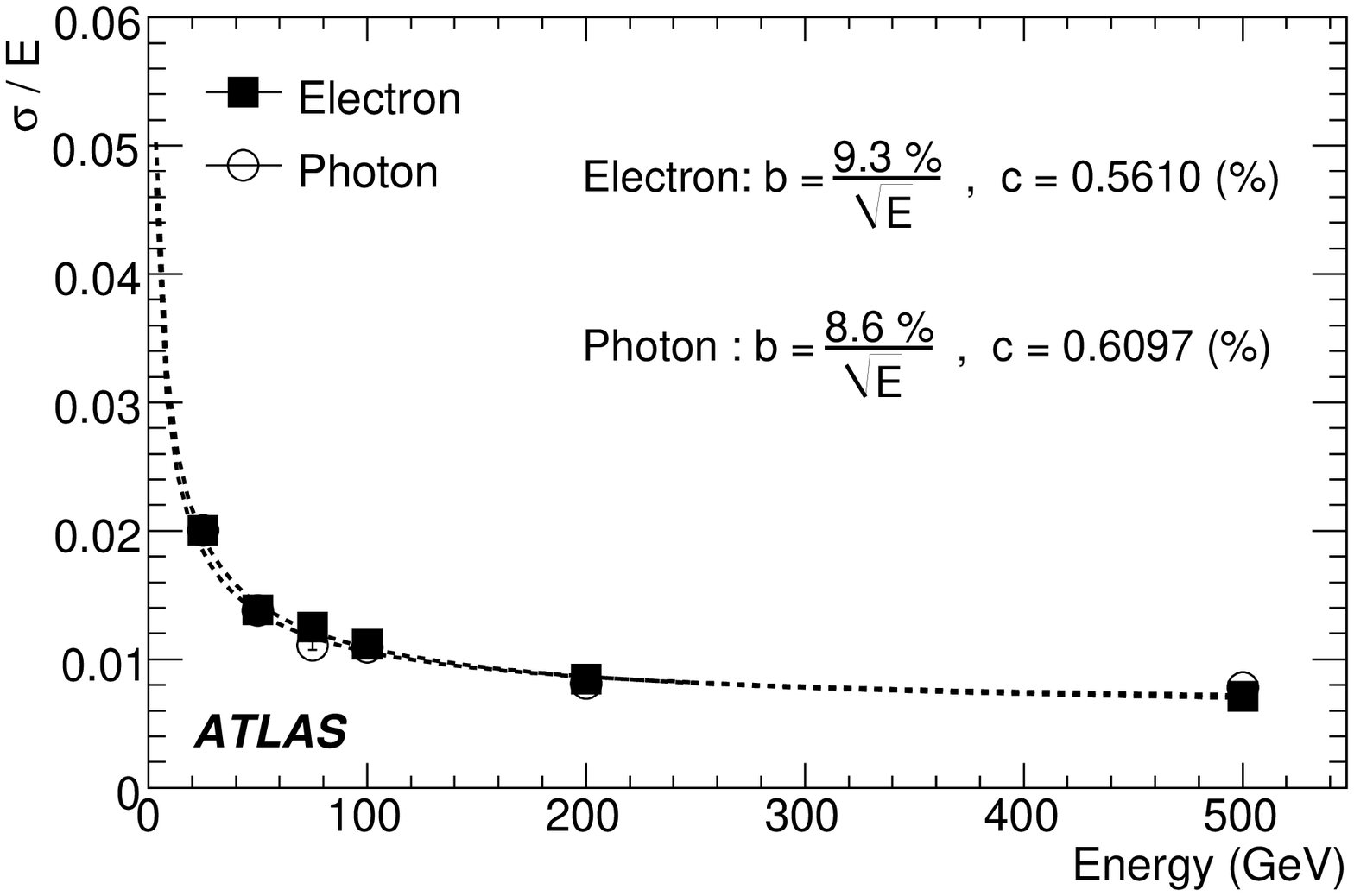}
\includegraphics[width=65mm]{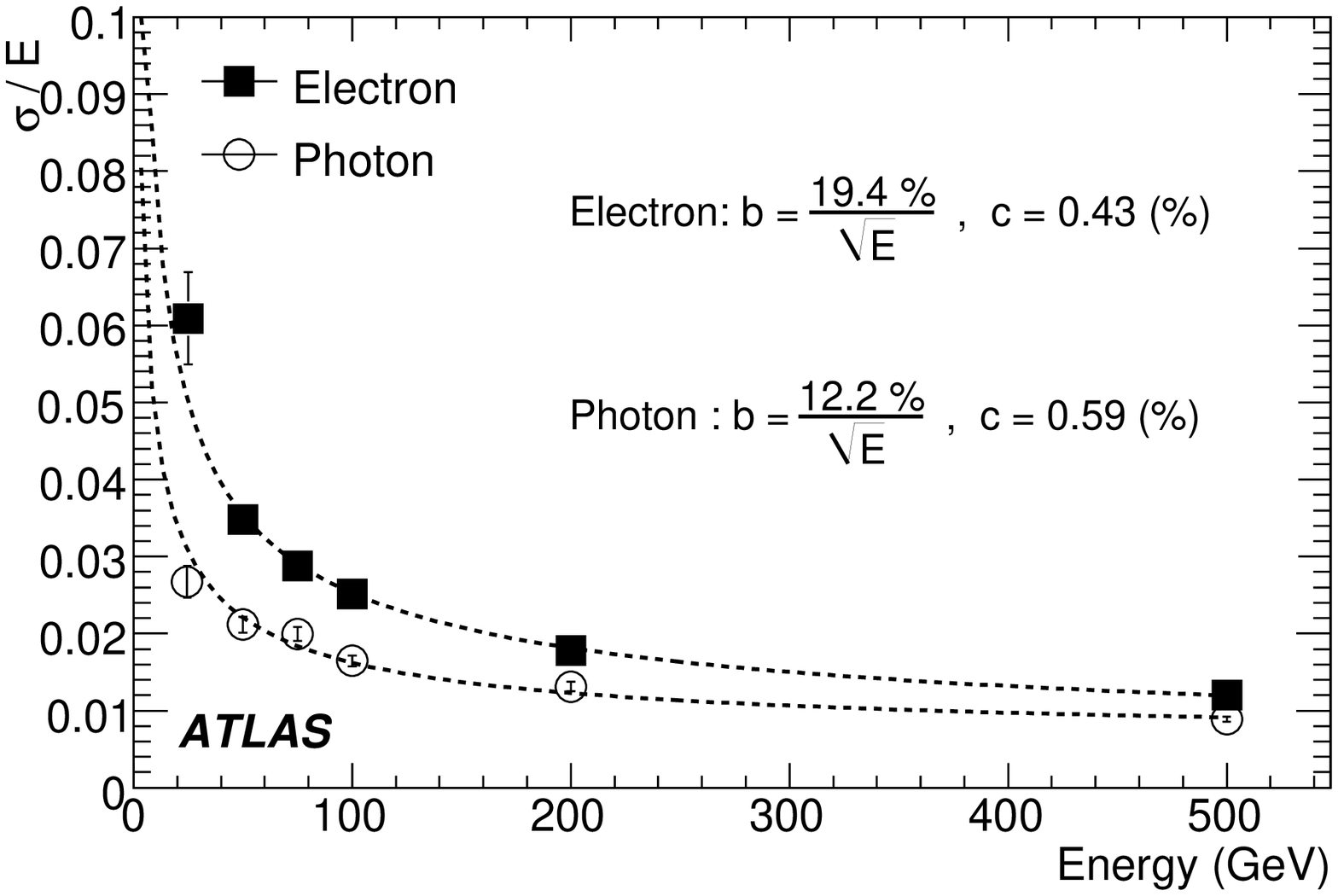}
\caption{Energy resolution for electrons and photons at $|\eta| = 0.3$ and $|\eta| = 1.65$, as function of incoming energy. This is obtained by using simulated single electron and photon samples.} \label{figure1}
\end{figure*}

Figure~\ref{figure1} shows the resolution as a function of the particle energy for electrons and photons at $|\eta| = 0.3$ and $|\eta| = 1.65$. The fits shown allow the extraction of a sampling term of the order of $10\% /\sqrt {E[GeV]}$ and a small constant term~\cite{eg6}. This result is confirmed by the analysis of real test beam data~\cite{testbeam}.

\begin{figure*}[t]
\centering
\includegraphics[width=65mm]{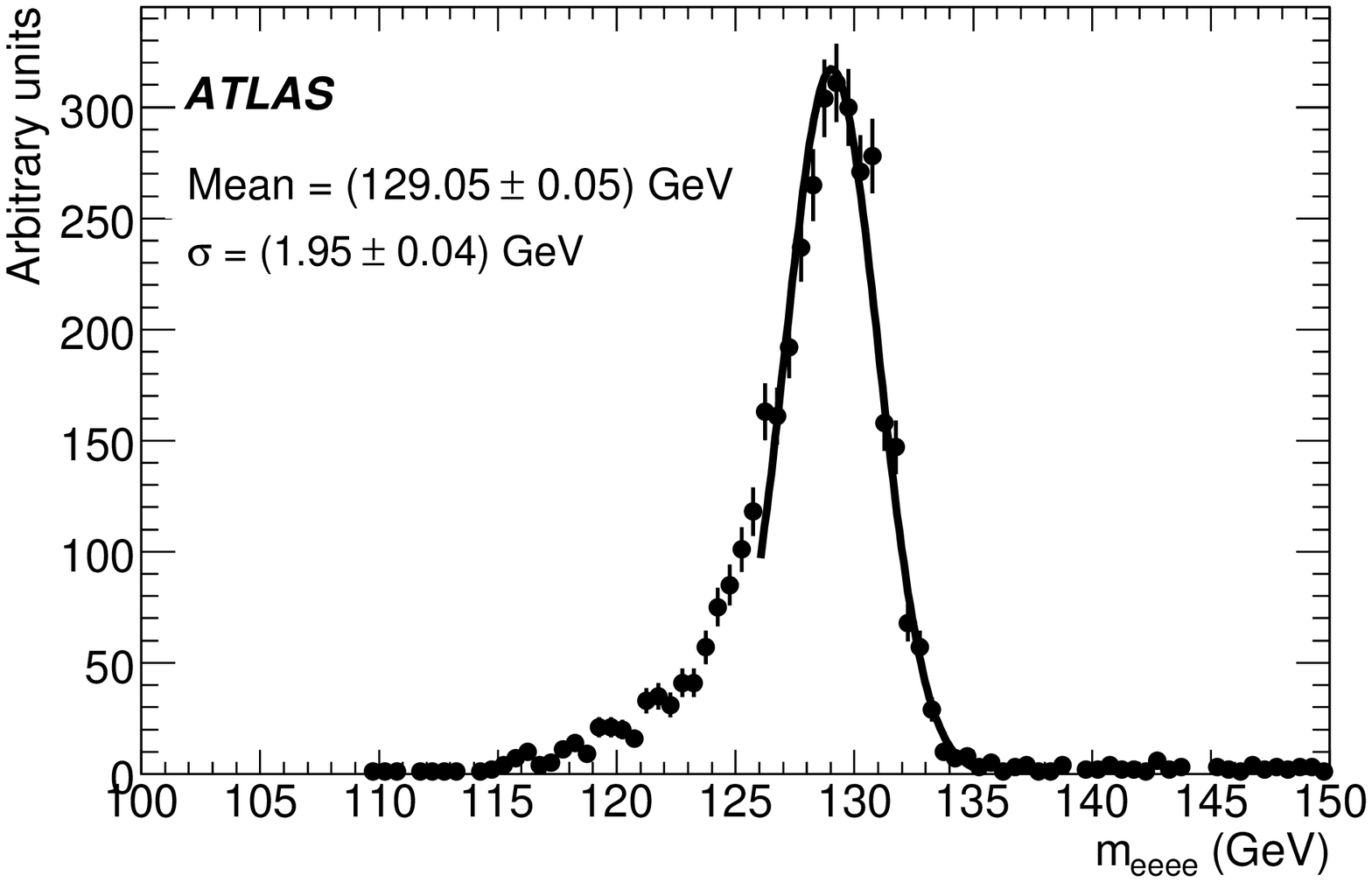}
\includegraphics[width=65mm]{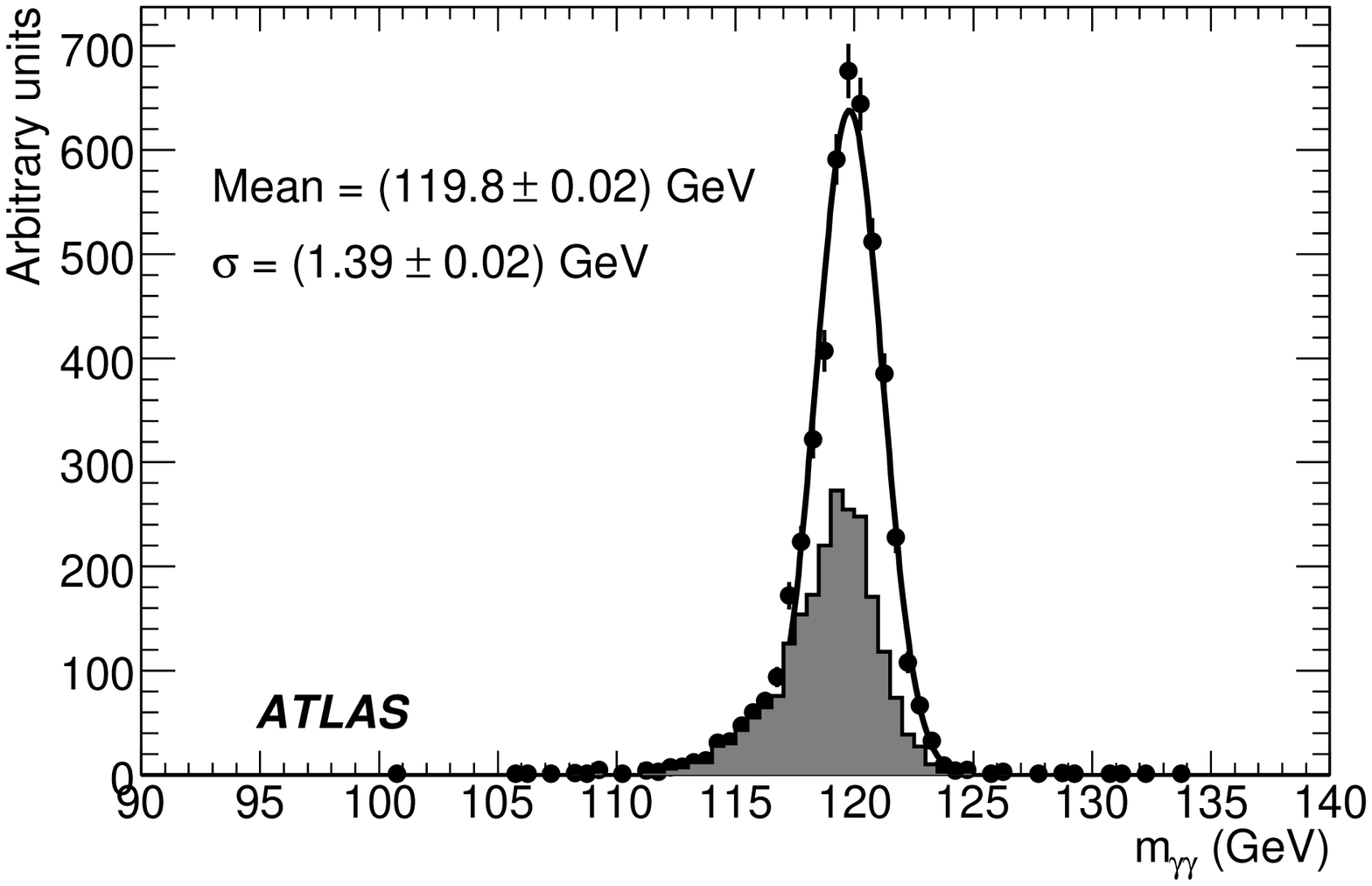}
\caption{(a) The invariant mass of four electrons ($m_{eeee}$) from Higgs boson decay samples with $m_{H} = 130$ GeV (using calorimetric information only, with no $Z$ boson mass constraint). (b) The invariant mass of two photons ($m_{\gamma \gamma}$) from Higgs boson decay with $m_{H} = 120$ GeV. The shaded plot corresponds to at least one photon converting at $r < 80 cm$.} \label{figure2}
\end{figure*}

Figure~\ref{figure2} (a) shows the reconstructed distribution of the invariant mass of the electrons after calibration, in the $H \to e e e e$ decay, with $m_{H} = 130$ GeV. The central value is correct at the $0.7\%$ level and with a Gaussian resolution of $1.5 \%$. Figure 2 (b) shows the reconstructed photon pair invariant mass for $H \to \gamma \gamma $ decays with $m_{H} = 120$ GeV. The central value of the reconstructed invariant mass is correct at $0.2\%$ level and with a Gaussian resolution of $1.2\%$~\cite{eg6}.   

Using the clean and large-statistics sample of $Z \to e e$, it is possible to evaluate the overall EM energy scale of the calorimeter from the data, and to determine precisely the inter calibration  between different regions of the calorimeter. Monte Carlo-based evaluations, using ~87,000 reconstructed $Z \to e e$ events, shows that the long-range constant term can be kept below $0.5\%$~\cite{eg6}. This gives a global constant term below the design value of $0.7 \%$.

\section{ELECTRON AND PHOTON RECONSTRUCTION}

The “sliding window” algorithm is used to find and reconstruct EM clusters. This forms rectangular seed clusters with a fixed size, $0.125 \times 0.125 ~ ( \eta \times \phi) $, positioned to maximize the amount of energy within the cluster. The combined reconstruction and classification checks whether a track can be matched to the seed cluster. If yes and the track does not correspond to a conversion, it is classified as an electron, else as a photon. The cluster is calibrated according to the particle hypothesis (electron/photon) with an optimized cluster size. 

Due to the structure of the ATLAS tracker, photons which convert within $300$ mm of the beam axis are associated with a track seeded in the silicon volume, while photons which convert further away from the beam pipe are found using tracks seeded in the Transition Radiation Tracker (TRT)~\cite{innertdr} with or without associated hits in the silicon detector volume~\cite{innertdr}. To reconstruct converted photon vertices, a dedicated vertex finder algorithm is used. Combining these tools, a reconstruction efficiency of almost $80 \%$ can be achieved for conversions that occur up to a distance of $800$ mm from the beam axis~\cite{eg4}.

Low momentum, so called soft, electrons from $J/\Psi$ and $\Upsilon$ decays are useful to determine in-situ performance of the trigger, offline reconstruction and to calibrate the EM calorimeter. For initial luminosities of $10^{31} ~cm^{-2} s^{-1}$, a trigger on low energy dielectron pairs (two Level 1 Trigger EM clusters greater than $3$ GeV) and tracking selection in the High Level Trigger should provide a large sample of soft electrons from direct production of $J/\Psi$ and $\Upsilon$. Track-seeded offline reconstruction of low energy electrons finds a track in the inner detector and extrapolates it to the middle layer of the EM calorimeter and apply energy and position corrections~\cite{eg6} to calorimeter EM cluster. With an integrated luminosity of $100 ~pb^{-1}$, a cut based electron identification and using the reconstruction of low-mass electron pairs, approximately two hundred thousand $J/\Psi$ decays could be isolated~\cite{eg7}.

\section{ELECTRON AND PHOTON IDENTIFICATION}

In order to separate real electrons and photons from jets, several discriminating variables are constructed by combining the information from the calorimeters and the inner tracking system. Calorimeter information is used to select events containing a high-$p_{T}$ EM shower. Track isolation is used to further reduce remaining fake photons from high-$p_{T} ~\pi_{0}$ low multiplicity jets. Electron indentification uses more sophisticated track information.

\begin{figure*}[t]
\centering
\includegraphics[width=75mm]{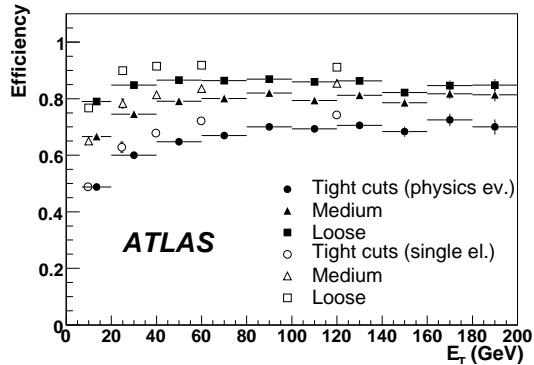}
\caption{Electron identification efficiency as a function of $E_{T}$. The full symbols correspond to electrons in SUSY events and the open ones to single electrons of fixed $E_{T}$. This figure is taken from Ref.~\cite{eg1}.} \label{figure3}
\end{figure*}

Cut-based identification of high $p_{T}$ electrons (photons) is based on many cuts which have been optimized in up to seven (six) bins in $\eta$ and up to six (eight) bins in $p_{T}$. Three levels of selections with increasing purity are available: loose, medium and tight~\cite{eg1}. Figure~\ref{figure3} shows the identification efficiency of the loose, medium and tight cuts as a function of $E_{T}$. The efficiencies are compared between single electrons of $E_{T} = 10, 25, 40, 60, 120 ~$GeV and electrons found in simulated supersymmetric events. As expected, the single electron sample displays higher efficiencies than in supersymmetric events, because of the large hadronic activity in the later type of events.

\begin{figure*}[t]
\begin{minipage}[b]{0.48\linewidth}
\centering
\includegraphics[width=65mm,height=45mm]{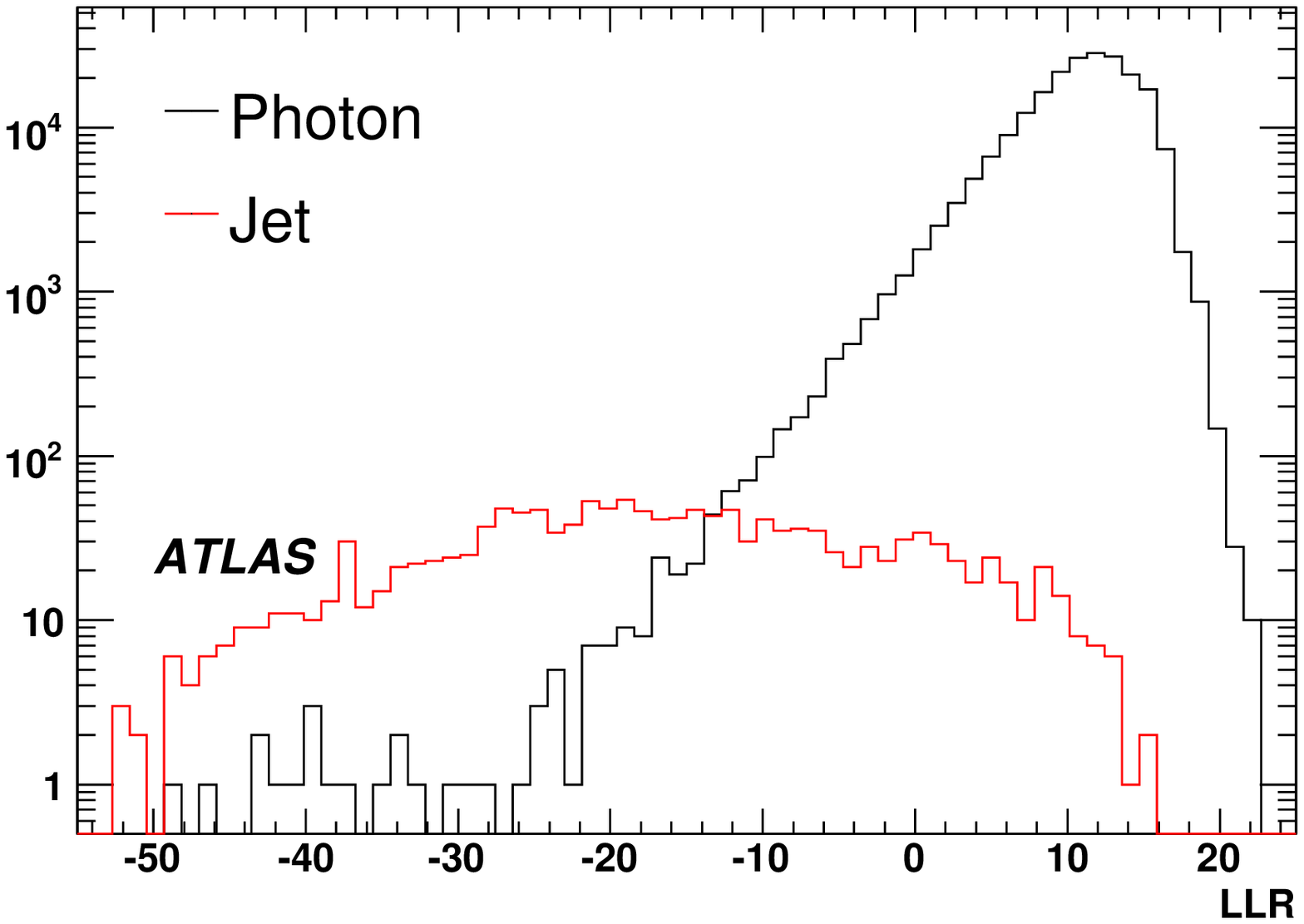}
\vspace{0.36cm}
\end{minipage}
\hspace{-2cm}
\begin{minipage}[b]{0.48\linewidth}
\centering
\vspace{0.4cm}
\includegraphics[scale=0.34]{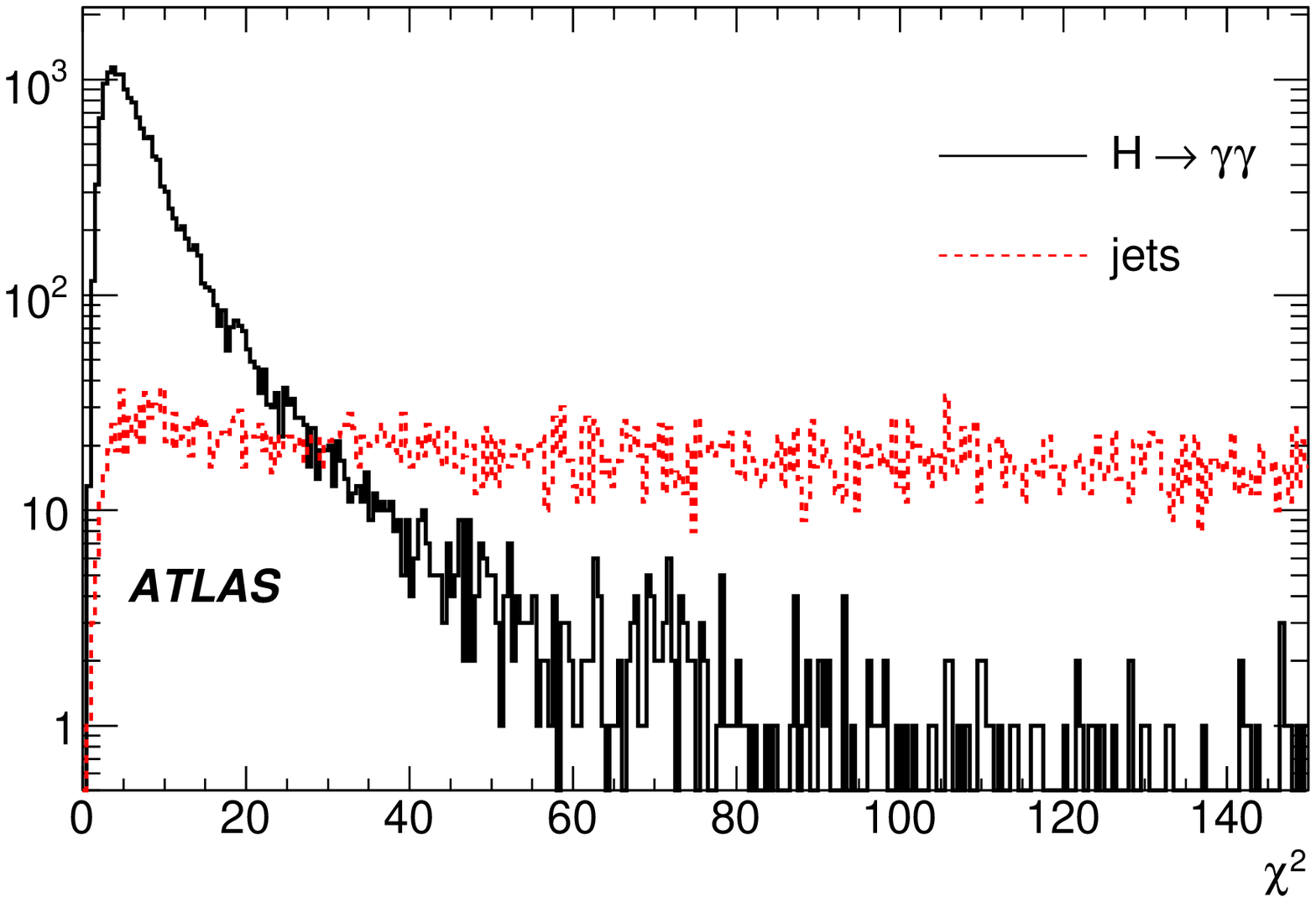} 
\vspace{0.08cm}
\end{minipage}
\vspace{-0.1cm}
\caption{(a) The distribution of LLR for photons (black histogram) and jets (gray histogram). (b) The H-matrix $\chi^2$ distribution for an inclusive jet sample (dashed histogram) and for the individual photons from the $H \to \gamma \gamma$ sample (solid histogram). These figures are taken from Ref.~\cite{eg2}.} \label{figure4}
\end{figure*}

In the Log-Likelihood Ratio (LLR) method, the distribution of each of the shower variables is normalized to unity to obtain a probability density function (PDF). Once the PDF's are established, the LLR value is computed as $ LLR  =  \sum_{i=1}^{n} \ln (Ls_{i} / Lb_{i})$, where $Ls_{i}$ and $Lb_{i}$ are PDF's of the $i^{th}$ shower shape variable for the real electrons/photons and the jets, respectively. Figure~\ref{figure4} (a) shows the distribution of LLR for photons and jets. The LLR cut can be tuned in bins of $\eta$ and $p_{T}$ to obtain an optimal separation between photons and jets.

The H-matrix method exploits the correlations among transverse and longitudinal shower shape variables to identify electrons and photons. The resemblance of a candidate to an electron or a photon shower is quantified by $\chi^2 = \Sigma^{\rm dim=10}_{i,j=1} ~ (y^{m}_{i} - \bar{y}_{i}) H_{ij} (y^{m}_{j} - \bar{y}_{j}) $, where $H= M^{-1}$ is the inverse of the covariance matrix $M$ of the shower shape variables, and the indices $i$ and $j$ run from 1 to the total number of variables, namely 10. The shape of the distributions of the selected shower shape variables depend on the $\eta$ and energy of the incoming photon or electron. These effects are taken into account in the construction of the H-matrix using single photon or electron samples of different energies, to parameterize each of the covariance terms in the matrix $M$ as a function of photon or electron energy. The separation power of the H-matrix between real photons and jets is illustrated in Figure~\ref{figure4} (b), where the $\chi^2$ distribution of the H-matrix for the jet samples is compared to that obtained for photons from the $H \to \gamma \gamma$ decay.

%%%%%%%%%%%%%%%%%%%%%%%%%%%%%%%%%%%%
%%%%%%%%%%%%%%%%%%%%%%%%%%%%%%%%%%%%

\begin{acknowledgments}
The author would like to thank ATLAS Electron/Photon working group.

\end{acknowledgments}


\begin{thebibliography}{9}   % Use for  1-9  references
%\begin{thebibliography}{99} % Use for 10-99 references

\bibitem{phytdr}
ATLAS collaboration, ATLAS Detector and Physics Performance Technical Design Report, CERN/LHCC 99-15, 1999.

\bibitem{caltdr}
ATLAS collaboration, Liquid Argon Calorimeter Technical Design Report, CERN/LHCC 96-41, 1996.

\bibitem{eg6}
ATLAS collaboration, ATLAS public note EG6, ``Electromagnetic Calorimeter Calibration and Performance'', 2008.

\bibitem{testbeam}
Aharrouche, M et al., Nucl. Instrum. Methods Phys. Res., A 582 (2007) 429-455.

\bibitem{innertdr}
ATLAS collaboration, Inner Detector Technical Design Report, CERN/LHCC 97-16/17, 1997.

\bibitem{eg4}
ATLAS collaboration, ATLAS public note EG4, ``Photon Conversion in ATLAS'', 2008.

\bibitem{eg7}
ATLAS collaboration, ATLAS public note EG7, ``Reconstruction of $J/\Psi \to e^{+} e^{-} $ and $\Upsilon \to e^{+} e^{-}$ Decays'', 2008.

\bibitem{eg1}
ATLAS collaboration, ATLAS public note EG1, ``Reconstruction and Identification of electrons in ATLAS'', 2008.

\bibitem{eg2}
ATLAS collaboration, ATLAS public note EG2, ``Expected Performance for the Reconstruction and Identification of Photons'', 2008.



%\bibitem{exampl-ref}
%A.N. Other, ``A Very Interesting Paper'', EPAC'96, Sitges, June
%1996.

\end{thebibliography}
\end{document}